\documentclass[twocolumn,preprintnumbers,amsmath,amssymb,superscriptaddress]{revtex4}

\usepackage{amsmath}
\usepackage{graphicx}

\begin{document}

\title{The Value of Conflict in Stable Social Networks}

\author{Pensri Pramukkul}
\affiliation{Center for Nonlinear Science, University of North Texas, P.O. Box 311427,
Denton, Texas 76203-1427, USA}
\affiliation{Faculty of Science and Technology, Chiang Mai Rajabhat University, 202 Chang
Phuak Road, Muang, Chiang Mai 50300, Thailand}

\author{Adam Svenkeson}
\affiliation{Center for Nonlinear Science, University of North Texas, P.O. Box 311427,
Denton, Texas 76203-1427, USA}
\affiliation{Army Research Laboratory, 2800 Powder Mill Road, Adelphi, MD 20783, USA}

\author{Bruce J. West}
\affiliation{Information Science Directorate, Army Research Office, Research Triangle
Park, NC 27709, USA}

\author{Paolo Grigolini}
\affiliation{Center for Nonlinear Science, University of North Texas, P.O. Box 311427,
Denton, Texas 76203-1427, USA}

\begin{abstract}
A cooperative network model of sociological interest is examined to
determine the sensitivity of the global dynamics to having a fraction of the
members behaving uncooperatively, that is, being in conflict with the
majority. We study a condition where in the absence of these uncooperative
individuals, the contrarians, the control parameter exceeds a critical value
and the network is frozen in a state of consensus. The network
dynamics change with variations in the percentage of contrarians, resulting
in a balance between the value of the control parameter and the percentage
of those in conflict with the majority. We show that the transmission of
information from a network $B$ to a network $A$, with a small fraction of
lookout members in $A$ who adopt the behavior of $B$, becomes maximal when
both networks are assigned the same critical percentage of contrarians.
\end{abstract}

\maketitle

The regulatory dynamics of the brain \cite{chialvo10}, the cardiovascular
and other physiological systems \cite{west06}, and indeed most
biological/sociological networks appear to be poised at criticality \cite{bialek}. The
existence of phase transitions is so common, in part, because criticality is
the most parsimonious way for a many-body system, with nonlinear
interactions to exert self-control. Inhibitory links in neurophysiology and
contrarians in sociology are the names given to interactions that evoke the
disruption of organization and consensus, thereby suggesting that the
well-being of either the brain or human society requires the containment of
those negative agents. However, recent neurophysiological literature shows
that this perspective may be overly restrictive, and that a sufficiently
large concentration of inhibitory links may counter-intuitively have the beneficial effect of promoting a ceaseless activity \cite{restrepo}, a characteristic that can provide evolutionary advantage. 

One of the first explanations of abrupt social transitions in terms of
criticality was made by Callen and Shapero in 1974 \cite{callen74}. They put
together the concepts of social imitation and critical behavior a generation
before Gladwell popularized the concept of the tipping point \cite%
{gladwell00}. In the sociological phenomenon of interest to us here the role
of inhibitory links is played by individuals called \emph{contrarians} \cite%
{galam}. More recently an interesting connection has been established
between the action of contrarians and the property of frustration found in spin glasses \cite{strogatz,germangirl}. As the term
frustration suggests, the action of contrarians is found to quench consensus
or prevent its occurrence in accordance with the sociological conclusions of
Crokidakis \textit{et al}. \cite{celia}. However, we reach a
different conclusion and find value in those individuals whose method of
decision making are in conflict with the majority.

Herein the observation made in neurophysiology \cite{restrepo} is adapted to
sociology using the decision making model (DMM). The complex network
described by the DMM implements the echo response hypothesis, which
assumes that the dynamic properties of a network of identical individuals is
determined by individuals imperfectly copying the behavior of one another 
\cite{book}. The effect of introducing contrarians into a cooperative social network is
analyzed using a system of coupled two-state master equations. Using analytical
calculations we show that in the presence of contrarians, increased cooperation
effort, in the form of increased values of the DMM control parameter, is
necessary to achieve consensus. At the same time, contrarians may promote a condition of ceaseless activity similar to that found in the cognitive context \cite{restrepo}. The maintenance of such a social state of alertness supports a kind of flexibility that has survival
value. Recall the well-known phrase, often attributed to Thomas Jefferson,
``Eternal vigilance is the price of liberty."

With the growing evidence of phase transitions in biological and sociological systems, recent studies have turned to information-theoretic analyses of canonical models for more insight. For instance, Refs. \cite{matsuda,gu,wicks,ribeiro} show that mutual information between elements peaks at the phase transition, while Ref. \cite{barnett13} shows that in the Ising model information flow between elements peaks in the disordered regime.  Following a different but related approach, we study the transmission of information from one DMM network to another through cross-correlation measurements.  We demonstrate that information transfer is maximally efficient when both systems are in a critical condition. This result is in accordance with earlier findings \cite{vanni,lukovic}, although here we point out that a non-vanishing fraction of contrarians is responsible for bringing the system to the critical point; information transfer would be negligible without those individuals in conflict with the majority present to disrupt the order in the system.    

DMM network dynamics is a member of the Ising universality class and is
found to be useful for describing phenomena related to social group behavior 
\cite{turalska,book}. The network model is based on the dynamics of single
individuals selecting one of two options. Denoting the two options as $+1$
and $-1$, the $i$-th individual generates the stochastic time series $%
s^{\left( i\right) }(t)=\pm 1$. Using the Gibbs perspective this time series
is analyzed using the solution to the two-state master equation 
\begin{align}
\frac{d}{dt}p_{1}^{(i)}(t)& =-\frac{1}{2}g_{12}^{(i)}(t)p_{1}^{(i)}(t)+\frac{%
1}{2}g_{21}^{(i)}(t)p_{2}^{(i)}(t)  \notag \\
\frac{d}{dt}p_{2}^{(i)}(t)& =-\frac{1}{2}g_{21}^{(i)}(t)p_{2}^{(i)}(t)+\frac{%
1}{2}g_{12}^{(i)}(t)p_{1}^{(i)}(t),
\end{align}%
where the $+1$ and $-1$ options have been labeled $1$ and $2$ respectively.
The time dependent transition rates $g_{12}^{(i)}(t)$ and $g_{21}^{(i)}(t)$
determine the production of decision events for the $i$-th individual. At
the moment of making a decision the single individual tosses a coin to
decide whether to keep the same opinion or to change her mind, hence the
factors of $1/2$ present in the master equation. Although this
implies a random decision, the interaction with the other individuals may
prolong or shorten the time necessary to make a decision, thereby generating
a bias toward one of the two choices.

For notational simplicity let us describe the behavior of the $i$-th
individual omitting the superscript $i$, while keeping in mind for now that
this is a single individual and that there are $N-1$ other individuals in
the network. The transition rate from state $|1\rangle $ to state $|2\rangle 
$ reads $g_{12}=g\exp \left[ -K\left( M_{1}-M_{2}\right) /M\right]$, where $%
M $ is the number of individuals linked to the $i$-th individual, $M_{1}$ is
the number of its neighbors in the state $|1\rangle $ and $M_{2}$ the number
of its neighbors in the state $|2\rangle $. When the interaction coupling
parameter $K$ vanishes, the $i$-th individual generates a Poisson sequence $%
s(t)$ with decision events generated at the fixed rate $g$. When $K>0$ the $%
i $-th individual cooperates with her neighbors. That is, when the $i$-th
individual is in the state $|1\rangle $ and the majority of her neighbors
share this state, the rate of her decision event productions decreases,
thereby indicating that the $i$-th individual is likely to remain in the
state $|1\rangle $ for a more extended time than in the absence of
interaction. The same cooperative prescription holds true when the $i$-th
individual is in the state $|2\rangle $, leading in this case to $%
g_{21}=g\exp \left[ -K\left( M_{2}-M_{1}\right) /M\right] $ and indicating
that if the majority of her neighbors are in the state $|1\rangle $ the $i$%
-th individual makes decisions with a faster rate, thereby reducing her
sojourn time in the state $|2\rangle $.

In this article we adopt the all-to-all (ATA) coupling condition, which
assigns to all the individuals the same number of neighbors, $M=N-1$, where $%
N$ denotes the total number of network members. Since the total number of
members usually satisfies $N\gg 1$, we set $M=N$. Under the ATA coupling
condition all the individuals are described by only two transition
rates.

The mean field of the network, 
\begin{equation}
\xi \left( t\right) =\frac{1}{N}\sum_{i=1}^{N}s^{\left( i\right) }(t)=\frac{%
N_{1}\left( t\right) -N_{2}\left( t\right) }{N},  \label{mean field}
\end{equation}%
becomes identical to a probability difference in the limit $N\rightarrow
\infty $ where $p_{i}=N_{i}/N$ ; $i=1,2$. Defining this probability
difference as $x\equiv p_{1}-p_{2},$ the master equation describing the mean
field behavior of an ATA DMM, consisting of an infinite number of
cooperative individuals, becomes 
\begin{equation}
\frac{dx}{dt}=\frac{g}{2}(e^{Kx}-e^{-Kx})-\frac{g}{2}(e^{-Kx}+e^{Kx})x.
\label{cooperators}
\end{equation}%
The equilibrium value of the mean field can be determined by setting the
left-hand side of Eq.~(\ref{cooperators}) equal to zero, which yields the
equation for the equilibrium value of the mean field: $x_{eq}=\tanh \left(
Kx_{eq}\right) .$ A second-order phase transition occurring in the cooperative
system at $K=1$ can be predicted as follows. If we make the assumption that
at the phase transition the equilibrium value of $x$ is very close to zero,
then using the Taylor series expansion of the hyperbolic tangent gives us $%
x_{eq}=Kx_{eq}$, which is compatible with a small but non-vanishing solution
only for $K=1$.

An individual is a contrarian if she is inclined to make a decision that is
the opposite of the one made by her neighbors \cite{galam}. Thus, for
instance, the $g_{12}$ transition rate would become $g_{12}=g\exp \left[
K\left( M_{1}-M_{2}\right) /M\right]$, and the $g_{21}$ transition rate
would become $g_{21}=g\exp\left[ -K\left( M_{1}-M_{2}\right) /M\right]$.
By following the same line of reasoning as that generating Eq.~(\ref%
{cooperators}) we obtain for an ATA DMM of contrarians 
\begin{equation}
\frac{dy}{dt}=\frac{g}{2}(e^{-Ky}-e^{Ky})-\frac{g}{2}(e^{Ky}+e^{-Ky})y.
\label{defectors}
\end{equation}%
We use the variable $y$ to denote the mean field of contrarians for the
purpose of distinguishing contrarians from cooperators. Recall that the
variable $x$ is associated with individuals that are cooperators.

When all the people in the network are contrarians, the network
remains close to the condition of a vanishing mean field. In fact, the
Taylor series expansion of the hyperbolic tangent in the solution now yields 
$y_{eq}=-Ky_{eq},$ which implies that the network remains fixed at the
equilibrium value $y_{eq}=0$, independently of the value $K$ of the
interaction strength.

\begin{figure}
\includegraphics{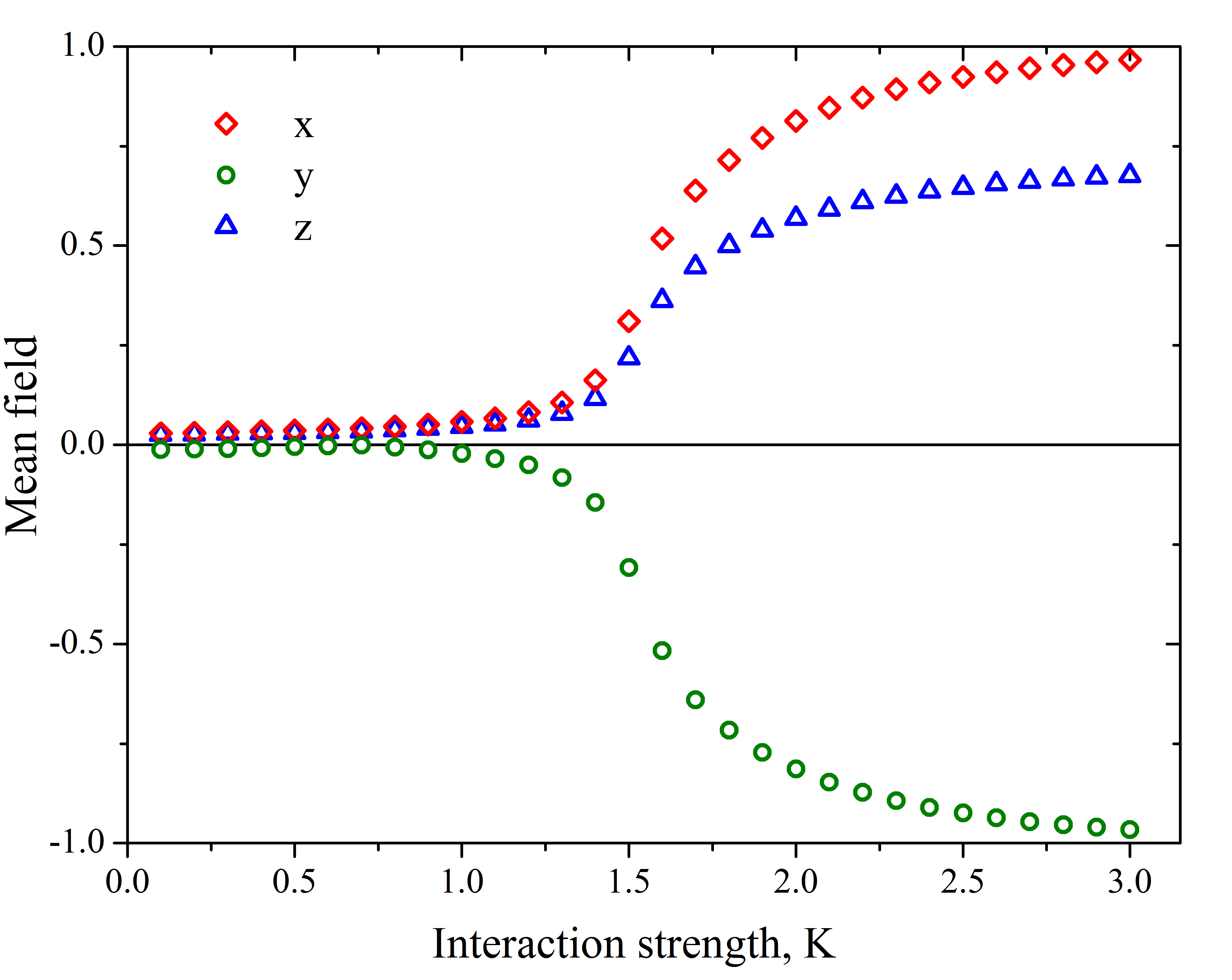}
\caption{Mean fields of the ATA DMM networks. The diamonds refer to the mean field of cooperators ($x$), the circles to the mean field of contrarians ($y$), and the triangles to the global mean field ($z$). $N = 10^3$ units, $g = 0.01$, and $q=0.15$. Note that the mean field of contrarians is opposite to the mean field of cooperators.}
\label{FIG1}
\end{figure}
We note however, see Fig.~\ref{FIG1}, that in a network with only a
small concentration of contrarians that a phase transition occurs with the
important symmetry property $y=-x,$ indicating that the mean field of
contrarians $y$ has the same intensity as the mean field of cooperators $x$,
but with the opposite sign. We are thus led to examining the condition where
the global field $z$, with a fraction $q$ of the individuals being
contrarians, is expressed by $z=(1-q)x+qy,$ which, using symmetry, becomes $%
z=(1-2q)x.$ The master equations for the cooperators can then be written 
\begin{equation}
\frac{dx}{dt}=g\sinh \left[ K(1-2q)x\right] -gx\cosh \left[ K(1-2q)x\right] ,
\label{DEFECTORSonCOOPERATORS}
\end{equation}%
and the master equation for the contrarians as 
\begin{equation}
\frac{dy}{dt}=-g\sinh \left[ K(1-2q)x\right] -gy\cosh \left[ K(1-2q)x\right]
.  \label{DEFECTORSonDEFECTORS}
\end{equation}

Note that the arguments of the exponential functions coincide with the
global field $z$, which is perceived by both cooperators and contrarians,
but contrarians react oppositely to that of the majority reaction to the
global field $z$. In fact, Eq.~(\ref{DEFECTORSonDEFECTORS}), determining the
time evolution of $y$, is obtained by replacing $K$ with $-K$ in the
exponential functions of Eq.~(\ref{DEFECTORSonCOOPERATORS}) representing the
influence of the neighbors on the decisions of single individuals.

We see that, as expected, the equilibrium condition generated by Eqs.~(\ref%
{DEFECTORSonCOOPERATORS}) and (\ref{DEFECTORSonDEFECTORS}) is $x_{eq}=\tanh
\left( K^{\prime }x_{eq}\right) =-y_{eq},$ where $K^{\prime }\equiv K(1-2q).$
It is evident that a phase transition occurs following the same mathematical
prescription as in the absence of contrarians with the main difference being
that the critical value of $K$ is now given by 
\begin{equation}
K_{c}(q)=\frac{1}{1-2q}.  \label{newcrucial}
\end{equation}%
Consequently, the critical control parameter increases in value as the
fraction of contrarians increases and the interaction effort necessary to
make a social decision in the presence of $50\%$ contrarians becomes
infinitely large. Therefore consensus cannot be reached beyond the limit of
50\% contrarians.

\begin{figure}
\includegraphics{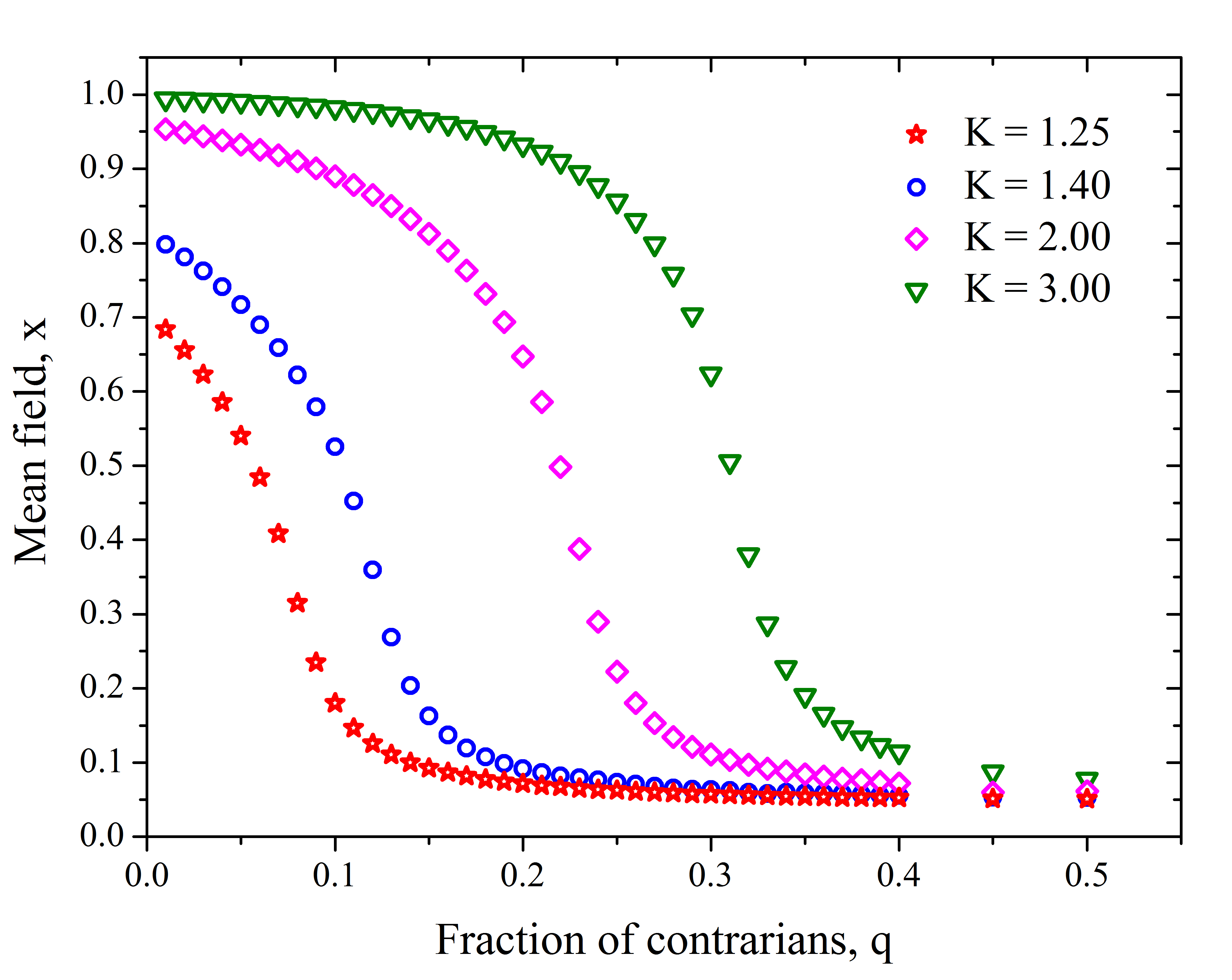}
\caption{The mean field of cooperators ($x$) with different fractions of contrarians ($q$) in the ATA DMM network of $10^3$ units with $g= 0.01$.  Increasing the fraction of contrarians has the effect of turning the supercritical into critical condition for a convenient fraction of contrarians. }
\label{FIG2}
\end{figure}
Fig.~\ref{FIG2} illustrates the formation of a social decision in a
diverse social network, having a mixture of cooperators and contrarians, and
the required strength of interaction coupling is in agreement with Eq.~(\ref%
{newcrucial}). The figure shows that when the network is in the
supercritical condition in the absence of contrarians, the action of an
increasing number of contrarians has the effect of shifting the network
dynamics down towards the critical point. Inverting Eq.~(\ref{newcrucial})
to obtain $q$ yields 
\begin{equation}
q_{c}(K)=\frac{1}{2}\left( 1-\frac{1}{K}\right) ,  \label{qc}
\end{equation}%
indicating that if the interaction strength $K$ is a fixed property of the
network, there exists a specific fraction of contrarians $q_{c}$ that will
bring the network to the critical point. Thus, the network dynamics can be
adjusted to operate at the critical point in three distinct ways: 1) with no
contrarians, a subcritical interaction strength can be increased to a
critical value; 2) with no contrarians, a supercritical interaction strength
can be decreased to a critical value and 3) for a fixed interaction strength
above the critical value, the fraction of contrarians can be increased to
the critical point for the network dynamics.

West \textit{et al}. \cite{west08} hypothesized that the maximum information
between complex networks occurs when the complexity of the two networks, as
measured by their respective inverse power law indices, match one another.
In the present context this hypothesis is tied to the criticality of the
network dynamics, since the time intervals between decision events display inverse power law behavior at criticality \cite{book}.
Moreover, the interesting discovery was made that the transmission of
information from a driving complex network $B$ to a driven network $A$
becomes maximally efficient when the two networks are at criticality \cite%
{vanni,lukovic}, that is, when they are maximally complex. Here we show that
when the percentage of contrarians induces criticality according to the
theory presented above, the condition of maximal efficiency for the transfer
of information is realized.

We consider two identical ATA DMM social networks each with $N=10^{3}$
individuals, with their interaction strength fixed at $K=1.25$. Note, that
Eq.~(\ref{qc}) predicts an interaction strength with 10\% contrarians is
necessary to realize criticality. To connect the driving network $B$ to the
driven network $A$, we introduce into network $A$ an additional 20
\textquotedblleft lookout\textquotedblright\ individuals that track the
global field of network $B$ \cite{sm}. This choice of
coupling between the two networks was inspired by a recent experiment \cite{nicolelis} where the signal from a few electrodes implanted in the brain of a rat $B$ is the information
transmitted directly to the brain of rat $A$, the whiskers of which, quite
surprisingly, synchronize with those of rat $B$.

Keeping an equal fraction of contrarians $q$ in both networks, we evaluate
the cross-correlation between network $A$ and network $B$ for various $q$
values \cite{sm}.
\begin{figure}
\includegraphics{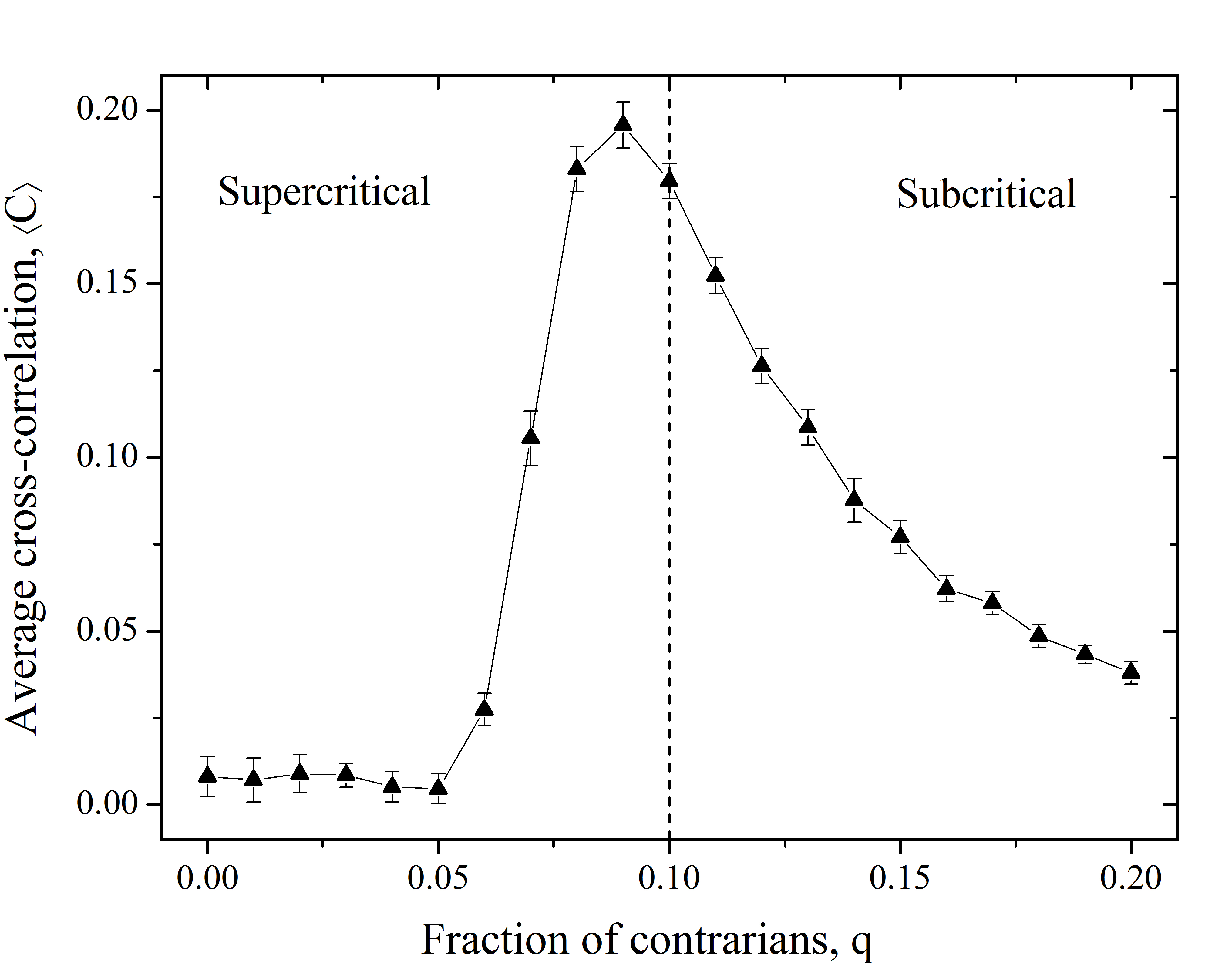}
\caption{The cross-correlation function ($\langle C\rangle$)between two predominantly cooperative networks (the ATA DMM networks with $g=0.01$ and $K = 1.25$) versus the fraction of contrarians ($q$). Each network shows a sharp peak in the vicinity of the predicted value.}
\label{FIG3}
\end{figure}
Fig.~\ref{FIG3} shows that the transmission of information from network $B$ to
network $A$ becomes maximally efficient at $q=0.09$. The numerical value for
the peak of the cross-correlation curve is near the theoretical value 
$q_{c}=0.1,$ predicted by Eq.~(\ref{qc}) on the basis of the conjecture that
criticality maximizes the efficiency of the information transport. One
possible reason for the slight discrepancy between the numerical results and
the value expected from theory is that Eq.~(\ref{qc}) does not account for
the presence of lookout individuals in the DMM network. A fraction of
lookout individuals $p$ creates the effective interaction strength 
$K_{p}=K(1-p)$ \cite{sm}. For the numerical
simulations we have $p=20/1020$ leading to $K_{p}\approx 1.225$. Replacing $K
$ by $K_{p}$ in Eq.~(\ref{qc}) yields $q_{c}=0.092$ for the critical
fraction of contrarians, refining the agreement with the results depicted
in Fig.~\ref{FIG3}.

We also see from Fig.~\ref{FIG3}, that when the concentration of
contrarians tends to vanish, the transmission of information between
networks becomes very small.  This is so because the social system falls in the supercritical condition,
which is not resilient and is unsuitable to address crucial issues. As a relevant example we have in mind the debate  on  setting a  limit to the extraction of energy resources 
so as not to leave the future generations empty handed \cite{nowak}. We believe that criticality corresponds to the condition of full democracy that according to the authors of Ref. \cite{nowak} would be necessary to promote
the energy sustainability of future generations, and we  imagine that this promotion may be realized  through a flow of information from one social network at criticality to another in the same democratic condition. 
On the other hand, a fraction of contrarians larger than the critical concentration $q_c$ of Eq. (\ref{qc}), realizing the subcritical condition,  is still compatible with 
a significant transmission of information, due to the distinctly asymmetric shape of $\langle C\rangle$ as a function of $q$.  Although our theoretical approach does consider information flow in the same manner as the recent work of Barnett 
\textit{et al}. \cite{barnett13},  it shows that the contrarian-induced disorder favors information transmission. 

While the results of the present article lend support to the attractive
discovery of the neural benefits of inhibitory links in cognitive networks 
\cite{restrepo}, our findings have a distinct sociological significance as
well. They suggest a kind of equivalence between two apparently quite
different forms of complexity, one of neurophysiological interest \cite
{restrepo} and the other of sociological interest, the latter pertaining to
the ATA DMM used herein.

The peaking of the cross-correlation function in Fig.~\ref{FIG3}
indicates that the concentration of contrarians within a network can be used
to establish a form of resonance between a driven and a driving network, a
central result of this paper. When the concentration $q$ is assigned the
critical value forcing the network to transition from a disordered state to
the condition when consensus is possible, the two networks establish a kind
of synchronization. 

The intensity of the cross-correlation becomes
negligible for values of $q$ smaller than the critical value, with the network freezing in a rigid consensus with a locked-in dependence among individuals, this being a condition of flawed democracy \cite{nowak}. Making $q$ larger than the critical value in the limiting case of a very large concentration of contrarians has the effect of allowing the individuals to recover their Poisson behavior, signaling the statistical independence between individuals. 
However, for an extended range of concentration of contrarians exceeding the critical value, the single units while recovering their freedom are still sensitive to criticality and  a significant amount of information transmission is possible.

We restricted the discussion of diversity to the mixture of cooperators and
contrarians in its simplest form, the ATA condition, using a mean field
approach. Moving away from the ATA condition, to a lattice
or generic network, introduces significantly
different global behavior. Specifically, the role of frustration must be
considered, as reaching consensus depends also on the topology of the distribution of contrarians with some configurations favoring and others obstructing criticality.  This more complicated scenario could be analyzed using a heterogeneous mean field theory approach.

\section*{Acknowledgements}
P.P., A.S., and P.G. warmly thank ARO and Welch for their support through Grants No. W911NF-11-1-0478 and No. B-1577, respectively.

\end{document}


\title{The Value of Conflict in Stable Social Networks: Supplementary Material}

\author{Pensri Pramukkul}
\affiliation{Center for Nonlinear Science, University of North Texas, P.O. Box 311427,
Denton, Texas 76203-1427, USA}
\affiliation{Faculty of Science and Technology, Chiang Mai Rajabhat University, 202 Chang
Phuak Road, Muang, Chiang Mai 50300, Thailand}

\author{Adam Svenkeson}
\affiliation{Center for Nonlinear Science, University of North Texas, P.O. Box 311427,
Denton, Texas 76203-1427, USA}
\affiliation{Army Research Laboratory, 2800 Powder Mill Road, Adelphi, MD 20783, USA}

\author{Bruce J. West}
\affiliation{Information Science Directorate, Army Research Office, Research Triangle
Park, NC 27709, USA}

\author{Paolo Grigolini}
\affiliation{Center for Nonlinear Science, University of North Texas, P.O. Box 311427,
Denton, Texas 76203-1427, USA}

\maketitle

\section*{Cross-correlation procedure}

\label{App A} The process of calculating the cross-correlation involves
first generating the mean field trajectory of the driving system $B$, $\xi
_{B}(t)$, and then using $\xi _{B}(t)$ to generate the corresponding mean
field trajectory of the driven system $\xi _{A}(t)$. Note that each lookout
individual in system $A$ adopts the time series $\xi _{l}(t)=\mathnormal{sign%
}\left\{ \xi _{B}(t)\right\} $, influencing the rest of the individuals in
system $A$; but we do not count the lookout individuals when calculating $%
\xi _{A}(t)$. Since we are interested in fluctuations around equilibrium,
from the mean field trajectories we calculate the deviation from their
time-averaged mean values, $\tilde{X}(t)=|\xi (t)|-\overline{|\xi (t)|}$,
where the overline indicates a time-averaged quantity and $|\cdot |$ the
absolute value. Finally, the cross-correlation (with zero delay) is
expressed as 
\begin{equation}
C =\frac{\overline{\tilde{X}_{A}(t)\tilde{X}_{B}(t)}}{\left(
\,\overline{\tilde{X}_{A}(t)^{2}}\,\,\overline{\tilde{X}_{B}(t)^{2}}%
\,\right) ^{1/2}},
\end{equation}%
The time averages are calculated over a finite time $T=10^{7}$, resulting in
the cross-correlation $C$ being a fluctuating quantity. For
this reason, in Fig.~$3$ of the main text we plot $\langle C
\rangle $, where the ensemble average indicated by the brackets was taken
over 10 independent realizations of networks $A$ and $B$, and the error bars
represent the standard deviation in the ensemble. 

Although $C $ fluctuates, it is important to note that the observation time of $T=10^{7}$
is much longer than the time it takes an isolated all-to-all system of size $N=10^{3}$ to reach equilibrium~\cite{bologna}. 
In fact, according to Ref. \cite{beig} the time necessary to get equilibrium, $T_{eq}$,  is given by
\begin{equation}
T_{eq}= \frac{1}{1.4\sqrt{\left(\gamma D \right)}}.
\end{equation}
This theoretical prediction is related to the decision making model in the all-to-all condition by setting \cite{bologna}
\begin{equation}
\gamma = \frac{g}{3} 
\end{equation}
and
\begin{equation} \label{missing}
D \propto \frac{g^2}{N}. 
\end{equation}
With these values we obtain $T_{eq} \approx 4\times10^4$. Note that the exact  proportionality factor of Eq. (\ref{missing}) is unknown, thereby making this estimate 
of $T_{eq}$ reliable only as far as the order of magnitude is concerned. This leads us to conclude that the adopted value of $T$ is much larger than $T_{eq}$ and that, 
consequently, we are
operating the network dynamics in the ergodic regime where the statistical
validity of time averages is ensured. The theory of this paper can be extended to the case $T < T_{eq}$, where important non-ergodic effects have to be taken into account, but this is left as
a subject of future investigation. 


\section*{Influence of lookout individuals on the critical point}

Here we discuss the influence of the lookout
individuals on the location of the critical point. When there are $l$
lookout individuals, the mean field of the complete network is 
\begin{equation}
\xi =\frac{N_{1}-N_{2}}{N}=\frac{n_{1}-n_{2}+l\,\xi _{l}}{N},
\end{equation}%
where $n_{1}$ and $n_{2}$ are the number of ordinary individuals in states $%
|1\rangle $ and $|2\rangle $, and $\xi _{l}$ is the state that the lookout
individuals adopt. Introducing $p=l/N$ for the fraction of lookout individuals, the mean
field becomes $\xi =(1-p)z+p\xi _{l},$ where we retain the notion that $z$
is the mean field of the ordinary individuals, consisting of a generic
mixture of cooperators and contrarians. Recalling that $z=(1-q)x+qy$, we
have that 
\begin{equation}
\frac{dz}{dt}=(1-q)\frac{dx}{dt}+q\frac{dy}{dt}.  \label{dzdt}
\end{equation}%
Going back to Eq.~(5) and Eq.~(6) of the main text, in the arguments of the exponential functions we
replace $z=(1-2q)x$ with $\xi $ for a diverse network, because the
transition rates of the ordinary individuals now depend on the mean field of
the full network, with lookout individuals included. We substitute the
result into Eq.~(\ref{dzdt}) to obtain the master equation 
\begin{equation}
\frac{dz}{dt}=(1-2q)g\sinh (K\xi )-gz\cosh (K\xi )
\end{equation}%
describing the mean field behavior of the ordinary units in the presence of
lookout individuals. The equilibrium mean field is defined by the equation 
\begin{equation}
z=(1-2q)\tanh (K\xi )\approx (1-2q)\tanh \left( K(1-p)z\right) ,
\end{equation}%
where the approximation ignores the contribution to $\xi $ from $p\xi _{l},$
under the assumption that there are only a few lookout individuals so that $%
p\ll 1$. Evaluating the value of the interaction strength at the critical
point by a Taylor expansion gives 
\begin{equation}
K_{c}(q,p)=\frac{1}{1-p}\frac{1}{1-2q}.
\end{equation}%
Alternatively, inverting the expression to obtain the fraction of
contrarians yields 
\begin{equation}
q_{c}(K,p)=\frac{1}{2}\left( 1-\frac{1}{K(1-p)}\right) .
\end{equation}%
Looking back to Eq.~(8) of the main text, we see that the effects of introducing
lookout individuals into the network can be accounted for by redefining an
effective interaction strength $K_{p}=K(1-p)$ among the ordinary people.